\documentstyle[aps,epsfig]{revtex}
\newcommand{\be}{\begin{equation}}
\newcommand{\ee}{\end{equation}}
\newcommand{\bea}{\begin{eqnarray}}
\newcommand{\eea}{\end{eqnarray}}
\newcommand{\X}{{\vec X}}
\newcommand{\pro}{\partial}
\newcommand{\n}{\hat n}
\newcommand{\oneg}{\displaystyle\frac{1}{g}}

\newcommand{\D}{{\hat D}}

\newcommand{\W}{{\vec W}}
\newcommand{\vX}{{\vec X}}

\newcommand{\vA}{{\vec A}}
\newcommand{\A}{{\vec A}}
\newcommand{\valpha}{{\vec \alpha}}

\newcommand{\hA}{{\hat A}}

\newcommand{\hD}{{\hat D}}
\newcommand{\hn}{\hat n}

\newcommand{\tC}{{\tilde C}}

\newcommand{\dfrac}{\displaystyle\frac}
\newcommand{\ba}{\begin{array}}
\newcommand{\ea}{\end{array}}

\newcommand{\nn}{\nonumber}

\begin{document}

\title{%
Extended QCD versus Skyrme-Faddeev Theory}

\author{
W. S. Bae, Y.M. Cho, and S. W. Kimm\\
{\it Department of Physics, College of Natural Sciences,
Seoul National University\\
Seoul 151-742, Korea\\
ymcho@yongmin.snu.ac.kr}}

\maketitle

\section*{Abstract}

We discuss the physical impacts of the ``Cho decomposition''
(or the ``Cho-Faddeev-Niemi-Shabanov decomposition'') of the
non-Abelian gauge potential on QCD. We show how the decomposition 
makes a subtle but important modification on 
the non-Abelian dynamics, and present three physically equivalent
quantization schemes of QCD which are consistent with the decomposition.
In particular, we show that the decomposition enlarges the 
dynamical degrees of QCD by making the topological degrees
of the non-Abelian gauge symmetry dynamical.
Furthermore with the decomposition we show that 
the Skyrme-Faddeev theory of non-linear
sigma model and QCD have almost identical topological structures. 
In specific we show that an essential
ingredient in both theories is the Wu-Yang type non-Abelian monopole,
and that the Faddeev-Niemi knots 
of the Skyrme-Faddeev theory can actually be interpreted to
describe the multiple vacua of the $SU(2)$ QCD. Finally we
argue that the Skyrme-Faddeev theory is, just like QCD, a theory 
of confinement which confines the magnetic flux of the monopoles.

\vspace{0.3cm}
PACS numbers: 12.38.-t, 11.15.-q, 12.38.Aw, 11.10.Lm

\section{Introduction}

Recently Faddeev and Niemi have discovered
three dimensional knot solutions
in the Skyrme-Faddeev theory of non-linear sigma model~\cite{faddeev1},
and made an interesting conjecture
that the Skyrme-Faddeev action could be interpreted as an effective
action for QCD in the low energy limit~\cite{faddeev2}.
On the other hand, in the low energy limit QCD is believed
to generate the confinement of color \cite{nambu}, which
is supposed to take place through the monopole
condensation \cite{cho1,cho2}. This suggests that
the Faddeev-Niemi conjecture and the confinement problem
can not be discussed separately. To prove the Faddeev-Niemi conjecture
one must construct the effective action
of QCD from the first
principles and produce the mass scale that the Skyrme-Faddeev
action contains, which could demonstrate the dynamical symmetry breaking
through the monopole condensation and triggers the confinement of color in
QCD. This has been very difficult to do. Fortunately
several authors have recently been able to argue that one can indeed
derive a ``generalized'' Skyrme-Faddeev action
from the effective action of QCD in
the infra-red limit, at least in some approximation \cite{lang,shab,gies}.
In fact there appears now a strong indication that the effective action
of QCD does generate a monopole condensation, and a generalized
Skyrme-Faddeev action could emerge as a crude approximation
from the effective action
near the vacuum condensation \cite{cho3}.
This strongly indicates that the Skyrme-Faddev action
is closely related to QCD in the infra-red limit.

In the discussion of the Faddeev-Niemi conjecture 
a gauge independent decomposition of
the non-Abelian connection into the dual and valence 
potentials \cite{cho1,cho2}
has played an important role \cite{lang,shab,gies}.
The purpose of this paper is  
to discuss the physical impacts of the
decomposition on QCD in more detail. To do this we first clarify
the considerable confusion and misunderstanding about the decomposition 
which has appeared in the literature recently \cite{faddeev2,shab,gies}. 
{\it With the clarification we show how the decomposition
makes a subtle but important modification of the non-Abelian
dynamics, and present three (physically equivalent) 
quantization schemes of QCD which are
consistent with the decomposition. In particular we show that 
the decomposition enlarges the dynamical degrees of QCD by making
the topological degrees of the non-Abelian gauge symmetry dynamical}.
This demonstrates that the decomposition has a deep impact
on the non-Abelian dynamics.

With this we show how the decomposition allows us 
to reveal almost identical topological structures which exist
between the Skyrme-Faddeev theory and QCD. Both
theories have interesting topological structures. For example, the
Skyrme-Faddeev theory has the topological knot solitons whose stability
comes from the topological quantum number. This is because the
non-linear sigma field of the knots defines the topological
mapping $\pi_3 (S^2)$ from the (compactified) three-dimensional
space $S^3$ to the internal space $S^2$. On the other hand it is
well-known that the non-Abelian gauge theory has a similar
topological structure. It allows infinitely many
topologically distinct vacua which one can label with an integer,
and that they are connected by the tunneling through the
instantons \cite{bpst,thooft}. In the $SU(2)$ gauge theory the
multiple vacua arise because the (time-independent) vacuum gauge
potential could be interpreted to define the topological mapping
$\pi_3(S^3)$ from the (compactified) 3-dimensional space $S^3$ to
the group space $S^3$, which could also be identified as the
mapping $\pi_3 (S^2)$ through the Hopf fibering \cite{cho4,white}. 
So one may wonder whether the topological
structures in two theories have any common ground. In this paper
we show that indeed the two theories have almost identical
topological structures. {\it With the decomposition 
we establish that an essential
ingredient in both theories is the Wu-Yang monopole, which
plays an important role in both theories. Furthermore we
demonstrate that the Faddeev-Niemi knots can actually be
identified to describe the multiple vacua of $SU(2)$ QCD}. 
With this observation we argue that the physical 
states of the Skyrme-faddeev theory 
are not the monopoles, but the knots. {\it This leads us to conjecture
that the Skyrme-Faddeev theory (just like QCD) is another
theory of confinement, where the non-linear self interaction of
the theory confines the magnetic flux of the monopole-anti-monopole
pairs to form the topologically stable knots}. This
confirms that the Skyrme-Faddeev non-linear sigma model is closely
related to QCD more than one way.

\section{Abelian Projection and Valence Potential}

Our discussion is based on the gauge independent decomposition of
the non-Abelian connection in terms of the restricted potential
(i.e., the dual potential) of the maximal Abelian subgroup $H$ of
the gauge group $G$ and the valence potential (i.e., the gauge
covariant vector field) of the remaining $G/H$ degrees
\cite{cho1,cho2}. Consider $SU(2)$ QCD for simplicity.  A natural
way to make the decomposition is to introduce an isotriplet unit
vector field $\n$ which selects the ``Abelian'' direction (i.e.,
the color direction) at each space-time point, and to
decompose the connection into the restricted potential (called the
Abelian projection) $\hat A_\mu$ which leaves $\n$ invariant and
the valence potential $\vec X_\mu$ which forms a covariant vector
field \cite{cho1,cho2},
\bea
& \vec A_\mu =A_\mu \hn - \oneg \hn
\times \pro_\mu\hn+\X_\mu
         = \hat A_\mu + \X_\mu, \nn\\
& A_\mu =\n\cdot \vec A_\mu,~~~~~~~\n^2 =1,
\eea 
where $A_\mu$ is the ``electric'' potential. Notice that
the restricted potential is precisely the connection which leaves
$\n$ invariant under the parallel transport, \bea \D_\mu \n =
\pro_\mu \n + g {\hat A}_\mu \times \n = 0. \eea Under the
infinitesimal gauge transformation \bea \delta \n = - \vec \alpha
\times \n \,,\,\,\,\, \delta \A_\mu = \oneg  D_\mu \vec \alpha,
\eea one has \bea &&\delta A_\mu = \oneg \n \cdot \pro_\mu
\valpha,\,\,\,\
\delta \hat A_\mu = \oneg \D_\mu \valpha  ,  \nn \\
&&\hspace{1.2cm}\delta \X_\mu = - \valpha \times \X_\mu  .
\eea
This shows that $\hat A_\mu$ by itself describes an $SU(2)$ connection which
enjoys the full $SU(2)$ gauge degrees of freedom. Furthermore
$\vec X_\mu$ transforms covariantly under the gauge transformation.
This confirms that our decomposition provides a gauge-independent
decomposition of the non-Abelian potential into the restricted part
$\hat A_\mu$ and the gauge covariant part $\vec X_\mu$.

We emphasize that the crucial element in our decomposition
(1) is the restricted potential $\hat A_\mu$. Once this part
is identified, the expression follows immediately due to the fact that the
connection space (the space of all gauge potentials) forms
an affine space \cite{cho1,cho2}. Indeed the affine nature of the
connection space guarantees that one can describe an arbitrary
potential simply by adding a gauge-covariant piece $\vec X_\mu$
to the restricted potential. Our decomposition (1), which has recently
become known as the ``Cho decomposition'' \cite{faddeev2}
or the ``Cho-Faddeev-Niemi-Shabanov
decomposition'' \cite{gies}, was
introduced long time ago in an attempt to demonstrate
the monopole condensation in QCD \cite{cho1,cho2}.
But only recently the importance of the decomposition
in clarifying the non-Abelian dynamics
has become appreciated by many authors \cite{faddeev2,lang}.
Indeed it is this decomposition which has played a crucial role to establish
the possible connection between the Skyrme-Faddeev action and the
effective action of QCD in the infra-red limit \cite{lang,shab,gies,cho3}, 
and the ``Abelian dominance''
in Wilson loops in QCD \cite{cho5}.

To understand the physical meaning of our decomposition notice that
the restricted potential $\hat{A}_\mu$ actually has a dual structure.
Indeed the field strength made of the restricted potential is decomposed as
\begin{eqnarray}
&\hat{F}_{\mu\nu}=(F_{\mu\nu}+ H_{\mu\nu})\hat{n}\mbox{,}\nonumber\\
&F_{\mu\nu}=\partial_\mu A_{\nu}-\partial_{\nu}A_\mu,
~~~~H_{\mu\nu}=-\dfrac{1}{g} \hat{n}\cdot(\partial_\mu\hat{n}\times\partial_\nu\hat{n})
=\partial_\mu \tilde C_\nu-\partial_\nu \tilde C_\mu,
\end{eqnarray}
where $\tilde C_\mu$ is the ``magnetic'' potential
\cite{cho1,cho2}. Notice that we can always introduce the magnetic
potential (at least locally section-wise), because
\bea
\partial_\mu {\tilde H}_{\mu\nu} = 0 ~~~ ( {\tilde H}_{\mu\nu} =
\dfrac{1}{2} \epsilon_{\mu\nu\rho\sigma} H_{\rho\sigma} ).
\eea
This allows us to  identify the non-Abelian monopole potential by
\bea
\vec C_\mu= -\frac{1}{g}\hat n \times \partial_\mu\hat n ,
\eea
in terms of which the magnetic field is expressed as
\bea
\vec H_{\mu\nu}=\partial_\mu \vec C_\nu-\partial_\nu \vec C_\mu+ g
\vec C_\mu \times \vec C_\nu =-\frac{1}{g}
\partial_\mu\hat{n}\times\partial_\nu\hat{n}=H_{\mu\nu}\hat n.
\eea

Another important feature of $\hat{A}_\mu$ is that, as an $SU(2)$
potential, it retains the full topological characteristics of the
original non-Abelian potential. Clearly the isolated singularities
of $\hat{n}$ defines $\pi_2(S^2)$ which describes the non-Abelian
monopoles.  Indeed $\hat A_\mu$ with $A_\mu =0$ and $\hat n= \hat
r$ (or equivalently, $\vec C_\mu$ with $\hat n= \hat r$) describes
precisely the Wu-Yang monopole \cite{wu,cho6}.  Besides, with the
$S^3$ compactification of $R^3$, $\hat{n}$ characterizes the Hopf
invariant $\pi_3(S^2)\simeq\pi_3(S^3)$ which describes the
topologically distinct vacua \cite{cho4,white}. This tells that
the restricted gauge theory made of $\hat A_\mu$ could describe
the dual dynamics which should play an essential role in $SU(2)$
QCD, which displays the full topological characters of the
non-Abelian gauge theory \cite{cho1,cho2}.

With (1) we have
\bea
\vec{F}_{\mu\nu}=\hat F_{\mu \nu} + \D _\mu \X_\nu -
\D_\nu \X_\mu + g\X_\mu \times \X_\nu,
\eea
and
\bea
\hn \cdot \X_\mu = 0, ~~~~~~~\hn \cdot \hat D_\mu \X_\nu = 0,
\eea
so that the Yang-Mills Lagrangian is expressed as
\bea
{\cal L}=&-&\dfrac{1}{4} \vec F^2_{\mu \nu }=-\dfrac{1}{4}
{\hat F}_{\mu\nu}^2 -\dfrac{1}{4} ( \D_\mu \X_\nu -
\D_\nu \X_\mu)^2 
-\dfrac{g}{2} {\hat F}_{\mu\nu}
\cdot (\X_\mu \times \X_\nu) \nn\\
&-& \dfrac{g^2}{4} (\X_\mu \times \X_\nu)^2
+ \lambda(\hat n^2 -1)
+ \lambda_\mu \hat n \cdot \X_\mu,
\eea
where $\lambda$ and $\lambda_\mu$ are the Lagrangian multipliers.
(Notice that, since $\vX_\mu$ is covariant, 
one might be tempted to put in a mass term
for the valence potential in the Lagrangian.
As we will see later, however, this is not allowed because the
mass term will spoil another type (the passive type) 
of gauge symmetry that the theory
should satisfy. In any case this issue becomes irrelevant when
the confinement sets in.) 
This shows that the Yang-Mills theory can be viewed as
the restricted gauge theory made of the dual potential $\hat A_\mu$,
which has
the valence gluon $\vec X_\mu$ as its source \cite{cho1,cho2}.

The equations of motion that one obtains
from our Lagrangian by varying $A_\mu$, $\vec X_\mu$,
and $\hat n$, are given by
\bea
&\partial_\mu ( F_{\mu\nu} + H_{\mu\nu} + X_{\mu\nu} ) = - g
~\hn \cdot [ \vX_\mu \times ( \hD_\mu \vX_\nu - \hD_\nu \vX_\mu )], \nn\\
&\hD_\mu ( \hD_\mu \vX_\nu - \hD_\nu \vX_\mu ) = g (
F_{\mu\nu} + H_{\mu\nu} + X_{\mu\nu} ) \hn \times \vX_\mu. 
\eea
where 
\bea 
X_{\mu\nu} = g ~\hn \cdot ( \vX_\mu \times \vX_\nu ).
\eea 
Observe that the equation of motion (12) is identical to 
\bea
D_\mu \vec F_{\mu\nu} = 0. 
\eea 
In fact with (1) one could have
obtained (12) directly from (14). This confirms that the decomposition
does not change the dynamics of QCD at the classical level.

\section{Re-interpretation of Non-Abelian Dynamics}

There is, however, a subtle but potentially disturbing
feature in our Lagrangian. The decomposition (1) contains $\hat
n$ explicitly, which seems to imply that here we are introducing
the $\hat n$ field as an extra variable which was absent
in QCD. In fact if one counts the independent degrees of our
decomposition, one finds that on the left-hand side one has 12
degrees before the gauge fixing. But on the right-hand side one
obviously has 14 [4 (for $A_\mu$) + 8 (for $\vec X_\mu$) + 2 (for
$\hat n$)] degrees. So it appears that with $\hat n$ we have ``two
extra degrees'' which we have to remove, if we could ever claim
that our Lagrangian indeed describes QCD. This observation has
created a considerable confusion in the literature,
and recently has led many people to search for
two extra constraints which can destroy the ``two
extra degrees'' created by $\hat n$ \cite{faddeev2,shab,gies}.

In fact, some authors
have advocated that one should impose 
the following extra constraint 
\bea
\hat D_\mu \vec X_\mu =0,
\eea
or a similar constraint on $\vec X_\mu$, to compensate 
the ``two extra degrees'' introduced by $\hn$ 
in our Lagrangian \cite{shab,gies}.
In the following we show that this is
based on the misunderstanding of our decomposition, and argue that
it is impossible to try to remove the ``two extra degrees'' imposing a
constraint on $\vec X_\mu$ by hand \cite{cho1,cho2}.
Indeed, the condition (15) is simply a necessary consistency condition 
which one need to remove the unphysical degrees
of $\vX_\mu$ to keep it massless, but it has nothing to do 
with the ``two extra degrees'' introduced by $\hn$.
First, to see whether our Lagrangian really has two extra degrees,
notice that the equations of motion (12) has no equation for $\hat n$
to satisfy. {\it Even though the Lagrangian (11) contains $\hat n$
explicitly, the variation with respect to $\hat n$ 
does not create any new equation of motion for $\hat n$
at all. This means that our $\hat n$ can not be interpreted as a
dynamical variable}. This, of course, is what one
should have expected. It must have been clear from (3) that our
$\hat n$ is a gauge artifact which could be removed completely
with a gauge transformation, at least locally section-wise. So,
even with $\hat n$ in the Lagrangian, we have no extra degrees to
remove, at least at the classical level. As for
the other 12 variables $A_\mu$ and $\vec X_\mu$, they are
obviously the legitimate dynamical variables of the $SU(2)$ gauge
theory, as the above equation of motion testifies. Secondly, the
fact that the connection space forms an affine space tells that
$\vec X_\mu$ must be completely arbitrary (except for the
constraint $\hat n \cdot \vec X_\mu=0$). {\it This means that by
adding any covariant $\X_\mu$ (which is orthogonal to $\hat n$) to
$\hat A_\mu$, we can always obtain a legitimate potential $\vec
A_\mu$}. In other word, putting a constraint on $\X_\mu$ amounts
to putting a constraint on $\A_\mu$ itself, which should only come
from a gauge fixing. 
Thirdly, the condition
(15) is nothing but the consistency condition for a vector field
theory. In fact, when $\vec X_\mu$ becomes massive, 
the constraint (15) actually follows
automatically from our equation of motion (12) as a consistency
condition. Indeed it
should be obvious that for any consistent theory of a vector field
one need this constraint, just like one need the well-known 5
constraints in the Pauli-Fierz equation of massive spin two field
for the consistency \cite{pauli}. Furthermore, 
when $\vec X_\mu$ remains massless, one has to impose a ``gauge
condition'' to supress the unphysical degrees of $\vec X_\mu$.
So any constraint on
$\vec X_\mu$ should come from a gauge fixing, and (15) should
be interpreted as a gauge condition which one need to
remove the unphysical degrees of $\vec X_\mu$. {\it Indeed 
it does become a gauge condition, when one tries
to quantize our Lagrangian with the background field method with
$\hat A_\mu$ as the background gauge potential, as we will show 
in the following}. Finally, it must be obvious from (4) that
the topological field $\hn$ becomes an essential ingredient of
the restricted potential $\hat A_\mu$, which has the full 
non-Abelian gauge degrees of freedom even without $\vec X_\mu$.
{\it So $\hn$ plays a fundamental role already in the restricted
gauge theory in the absence of $\vec X_\mu$,
and should be viewed completely independent of $\vec X_\mu$}. This
tells that it is futile to try to remove the ``two extra degrees''
created by $\hn$ from our Lagrangian by imposing the constraint (15), 
or any other constraint on $\vec X_\mu$, by hand.

Nevertheless it should also be evident that our Lagrangian has
indeed two extra degrees. To see this suppose we remove $\hat n$
completely with a gauge transformation. In this case $\hat n$
disappears, but it re-enters in a different form as the magnetic
potential \cite{cho1,cho2}, when (and only when) $\hat n$ contains
the topological singularities which define the topological quantum
number $\pi_2 (S^2)$. To demonstrate this, introduce the $SU(2)$
matrix element $S$ and parameterize it with $\alpha,~\beta$, and
$\gamma$,
\begin{eqnarray}
S = \exp{(-t_3\gamma)}\exp{(-t_2\alpha)}\exp{(-t_3\beta) },\nonumber
\end{eqnarray}
where $t_i$ are the adjoint representation of the $SU(2)$ generators,
and let
\begin{eqnarray}
&\hat{n}_i  = S^{-1} \hat{e}_i ~~~(i=1,2,3), \nn\\
&\hat e_1=(1,0,0),~\hat e_2=(0,1,0),~\hat e_3=(0,0,1), \nn\\
&\hat{n} =\hat{n}_3=(\sin{\alpha}\cos{\beta},~\sin{\alpha}\sin{\beta},~\cos{\alpha}).
\end{eqnarray}
Then under the gauge transformation $S$, one has
\begin{eqnarray}
\hat n &\longrightarrow& \hat e_3, \nn\\
\hat{A}_\mu &\longrightarrow& (A_\mu + \tilde C_\mu)\hat{e}_3,\nonumber\\
\quad \hat{F}_{\mu\nu} &\longrightarrow&(F_{\mu\nu}+H_{\mu\nu})\hat{e}_3,
\end{eqnarray}
where
\begin{eqnarray}\tilde C_\mu &=& \frac{1}{g} (\cos{\alpha}\partial_\mu\beta+
\partial_\mu\gamma) = -\frac{1}{g}\hat{n}_1\cdot\partial_\mu\hat{n}_2.
\end{eqnarray}
This shows that, in this magnetic gauge where $\hn$ disappears completely,
it is replaced by $\tilde C_\mu$ which describes precisely
the Dirac's monopole potential
around the isolated singularities of $\hat{n}$.
{\it This tells that our $\hat n$ describes the topological degrees,
not the local (i.e., dynamical) degrees, of
the non-Abelian gauge theory}. This is why the Lagrangian (11)
has no equation of
motion for $\hat n$.

The ``equation of motion'' for our $\hat n$ comes from an
unexpected quarter. Remember that we are dealing with a gauge
theory, and we have to impose a gauge condition. Since in our
decomposition $\vec X_\mu$ becomes a gauge covariant 
charged source, we can impose
a gauge condition to the restricted potential $\hat A_\mu$. We may
choose the following gauge condition, \bea
\partial_\mu \hA_\mu = ( \partial_\mu A_\mu ) \hn + A_\mu
\partial_\mu \hn - \dfrac{1}{g} \hn \times \partial^2 \hn = 0 .
\eea This could be re-expressed equivalently as two independent
conditions 
\bea
\partial_\mu A_\mu = 0 ,
\eea and \bea \hn \times \partial^2 \hn - g A_\mu \partial_\mu \hn
= 0. 
\eea
Now, observe that the last equation looks like a
perfect equation of motion for $\hat n$. {\it In fact we can
obtain exactly the same equation by varying $\hat n$,  if we
include a mass term for $\hat A_\mu$ in the Lagrangian (11)}. The
reason is that the mass term of $\hat A_\mu$ breaks the gauge
symmetry, so that one can no longer remove $\hat n$ with a gauge
transformation. This symmetry breaking generates the above equation
of motion for $\hn$, and makes it dynamical. {\it This tells that 
one can make the topological field $\hn$ dynamical by 
imposing a gauge condition}. This is not surprising, 
because we have already shown that in the magnetic gauge
where $\hn$ disappears completely, the topological field 
is transformed into the magnetic potential which can be 
treated as dynamical.

The lesson that we learn from the above discussion is
unmistakable. The constraint (15) is just a consistency condition
that one need to remove the unphysical degrees of $\vX_\mu$
to make it transverse and keep it massless,
which does not remove the extra degrees created by $\hn$ at all. Furthermore, 
the topological field $\hat n$ becomes dynamical
with the gauge fixing. 
This raises a very interesting question: How
many degrees do we have in $SU(2)$ QCD? The popular wisdom tells
us that we have 6 degrees of freedom after the gauge fixing. {\it But
with our decomposition (1) we have two more degrees after the
gauge fixing, even with the constraint (15)}. 
This seems to suggest that our decomposition 
does modify QCD in a subtle but important way.
How can we implement this idea and modify QCD to make it 
consistent with our decomposition? 
To show how, we will discuss three different but equivalent
methods to quantize the Lagrangian (11)
with the decomposition in the following. 

\section{Quantization of Extended QCD}

Certainly our decomposition leaves the Lagrangian invariant under 
the gauge transformation (3). But notice that the decomposition
inevitably introduces a new type of gauge symmetry which is different
from the gauge symmetry described by (3). 
This is because, for a given $\vA_\mu$,
one can have infinitely many different decomposition of (1),
with different $\hat A_\mu$ and $\vec X_\mu$ by choosing 
different $\hn$. Equivalently, for a fixed $\hn$, one can have
infinitely many different $\vec A_\mu$ which are gauge-equivalent
to each other. So it must be clear that with our decomposition we 
automatically have another type of gauge invariance which
comes from different choices of decomposition, which one has to deal with
in the quantization.  
This type of extra gauge degree of freedom is what one encounters 
in the background field method \cite{dewitt,pesk}. 
In the background field method
one decomposes the field into two parts, the slow-varying classical
part and the fluctuating quantum part, and deals with two type of
gauge transformations, the one acting on the slow-varying fields
and the other acting on the fluctuating fields, to obtain the
effective action of the slow-varying fields. 
This doubling of the gauge
degrees of freedom, of course, is an unavoidable consequence of 
the arbitrariness of the decomposition. In our case we can
adopt a similar attitude and quantize the Lagrangian (11) in two steps,
treating $\hat A_\mu$ and $\vec X_\mu$ separately. 

With this strategy we introduce two types of 
gauge transformations, the active (background)
gauge transformation and the passive (quantum) gauge transformation.
Naturally we identify (3) as the active gauge transformation.
As for the passive gauge transformation we choose
\bea
\delta \hn = 0, ~~~~~~~\delta \vec A_\mu = \dfrac{1}{g} D_\mu \vec \alpha,
\eea
under which we have
\bea
&\delta A_\mu = \dfrac{1}{g} \hn \cdot D_\mu \vec \alpha,
~~~~~~~\delta \hat A_\mu = \dfrac{1}{g} (\hn \cdot D_\mu \vec \alpha) \hn, \nn\\
&\delta \vec X_\mu = \dfrac{1}{g} [D_\mu \vec \alpha
-(\hn \cdot D_\mu \vec \alpha) \hn].
\eea
With this we can quantize $\vec X_\mu$ first by fixing 
the passive gauge degrees of freedom. We can definitely choose (15) 
as the gauge fixing condition for 
the passive type gauge transformation (23), because  
under (23) the constraint (15) no longer transforms covariantly.
In this case 
the Faddeev-Popov determinant corresponding to the gauge
condition (15) is given by
\bea
K_{ab} = \dfrac{\delta (\hat D_\mu \vec X_\mu)_a}{\delta \alpha^b} 
= [(\hat D_\mu D_\mu)_{ab} - n_a (\hn \cdot \hat D_\mu D_\mu)_b
+ g(\hn \times \vec X_\mu)_a (\hn \cdot D_\mu)_b].
\eea
The determinant has a remarkable feature, 
\bea
n_a K_{ab} = K_{ab} n_b =0.
\eea
This is because the determinant (24)
must be consistent with the constraints (2) and (10),
so that we must have
\bea
\delta (\hn \cdot \hat D_\mu \X_\mu) = \hn \cdot 
\delta (\hat D_\mu \X_\mu) = 0, \nn\\
\hat D_\mu D_\mu \hn = 0, ~~~~~~~\hn \cdot D_\mu \hn = 0,
\eea 
when $\hat D_\mu \vec X_\mu = 0$. This means that 
the ghost fields which correspond
to the determinant (24) should be orthogonal to $\hn$, and have only
two (not three) degrees of freedom.
With this gauge fixing of the passive type in mind, we now can make the
gauge fixing of the active type gauge transformation (3) with
the gauge condition (19). 
In this case the corresponding Faddeev-Popov determinant 
is given by
\bea
M_{ab} = \dfrac{\delta (\partial_\mu \hat A_\mu)_a}{\delta \alpha^b}
= (\partial_\mu \hat D_\mu)_{ab}.
\eea
From this we obtain the follwing generating functional
\bea
&W\{\hat j, j_\mu, \vec j_\mu\} = \dfrac{}{}\int {\cal D} \hn {\cal D}
A_\mu {\cal D} \X_\mu {\cal D} \vec{\eta_\perp} ~{\cal D}\vec{\eta_\perp}^{*}
{\cal D} \vec{c} ~{\cal D}\vec{c}^{~*} \exp[\dfrac{}{}
i\int{\cal L}_{eff} d^4x], \nn\\
&{\cal L}_{eff} =-\dfrac{1}{4}
{\hat F}_{\mu\nu}^2  
-\dfrac{1}{2\xi_1}(\partial_\mu \hat A_\mu)^2
+ {\vec c}^{~*}\partial_\mu \hat D_\mu \vec c +\lambda (\hn^2 - 1) \nn\\
&-\dfrac{1}{4} ( \D_\mu \X_\nu -
\D_\nu \X_\mu)^2 
-\dfrac{g}{2} {\hat F}_{\mu\nu}
\cdot (\X_\mu \times \X_\nu)
- \dfrac{g^2}{4} (\X_\mu \times \X_\nu)^2 \nn\\
&- \dfrac{1}{2 {\xi}_2} (\hD_\mu {\X}_\mu)^2 
+ {\vec \eta_\perp}^{~*}\hat D_\mu D_\mu \vec \eta_\perp 
-g^2({\vec \eta_\perp}^{~*} \times \X_\mu) \cdot (\X_\mu \times \vec \eta_\perp)
+ \lambda_\mu \hat n \cdot \X_\mu \nn\\
&+ \hn \cdot \hat j + A_\mu j_\mu + \X_\mu \cdot \vec j_\mu,
\eea
where $\vec j, j_\mu, \vec j_\mu$ are the external sources of
$\hn, A_\mu, \vec X_\mu$, and $\vec c$, ${\vec c}^{~*}$, 
$\vec \eta_\perp$, ${\vec \eta_\perp}^{~*}$ 
are the ghost fields of the active and passive gauge transformations
(Remember that here $\vec \eta_\perp$ and ${\vec \eta_\perp}^{~*}$
are orthogonal to $\hn$). 
This clearly shows how we can 
implement the consistency condition
(15) as a legitimate gauge condition. But we emphasize again
that in this quantization the topological field $\hn$ becomes 
a real dynamical field of QCD, even with the gauge fixing (15).
This is because, after the gauge fixing, the effective 
Lagrangian becomes a non-trivial function of $\hn$.

Another way to quantize the theory is by making
a slightly different decomposition. Remember that in our decomposition (1)
we have defined $A_\mu=\hn \cdot \vec A_\mu$, which gives the constraint
$\hn \cdot \vec X_\mu=0$. We can relax this condition by dividing
$A_\mu$ again into two parts, and let
\bea
&A_\mu = B_\mu + W_\mu, 
~~~~~~~\vec A_\mu = \check A_\mu + \W_\mu, \nn\\
& \check A_\mu = B_\mu \hn - \oneg \hn \times \pro_\mu \hn, 
~~~~~~~\W_\mu = \X_\mu + W_\mu \hn.
\eea
Notice that in this decomposition $\W_\mu$ need no longer be
orthogonal to $\hn$. Now, we can introduce the following 
active and passive gauge transformations for this decomposition
\bea
\delta \hn = -\vec \alpha \times \hn,
~~~~~~~\delta \check A_\mu = \dfrac{1}{g} \check D_\mu \vec \alpha,
~~~~~~~\delta \W_\mu = -\vec \alpha \times \W_\mu,
\eea
and
\bea
\delta \hn = 0, ~~~~~~~\delta \check A_\mu = 0, ~~~~~~~\delta \W_\mu 
=\dfrac{1}{g} D_\mu \vec \alpha.
\eea
Notice that both recover the original gauge transformation,
\bea
\delta \vec A_\mu = \dfrac{1}{g} D_\mu \vec \alpha.
\eea
With this we can quantize $\check A_\mu$ and $\W_\mu$ with the gauge conditions
\bea
\partial_\mu \check A_\mu = 0,
~~~~~~~\check D_\mu \W_\mu = 0.
\eea
In this case the corresponding Faddeev-Popov determinants 
are given by
\bea
M_{ab} = \dfrac{\delta (\partial_\mu \check A_\mu)_a}{\delta \alpha^b}
= (\partial_\mu \check D_\mu)_{ab}, 
~~~~~~~K_{ab} = \dfrac{\delta (\check D_\mu \W_\mu)_a}{\delta \alpha^b} 
= (\check D_\mu D_\mu)_{ab}.
\eea
From this we obtain the following generating functional
\bea
&W\{\hat j, j_\mu, \vec j_\mu\} 
= \dfrac{}{}\int {\cal D} \hn {\cal D}
B_\mu {\cal D} W_\mu {\cal D} \X_\mu {\cal D} 
\vec{\eta} ~{\cal D}\vec{\eta}^{~*}
{\cal D} \vec{c} ~{\cal D}\vec{c}^{~*} 
\delta (W_\mu-W^{(0)}_\mu)
\exp[\dfrac{}{}
i\int{\cal L}_{eff} d^4x], \nn\\
&{\cal L}_{eff} =-\dfrac{1}{4}
{\check F}_{\mu\nu}^2  
-\dfrac{1}{2\xi_1}(\partial_\mu \check A_\mu)^2
+ {\vec c}^{~*}\partial_\mu \check D_\mu \vec c + \lambda (\hn^2-1) \nn\\ 
&-\dfrac{1}{4} ( \check D_\mu \W_\nu -
\check D_\nu \W_\mu)^2 - \dfrac{1}{2} {\check F}_{\mu\nu} 
\cdot \check D_\mu \W_\nu 
-\dfrac{g}{2} \check D_\mu \W_\nu \cdot (\W_\mu \times \W_\nu) 
-\dfrac{g}{2} {\check F}_{\mu\nu}
\cdot (\W_\mu \times \W_\nu) \nn\\ 
&- \dfrac{g^2}{4} (\W_\mu \times \W_\nu)^2 
- \dfrac{1}{2 {\xi}_2} (\check D_\mu {\W}_\mu)^2
+ {\vec \eta}^{~*}\check D_\mu D_\mu \vec \eta 
+ \lambda_\mu \hat n \cdot \X_\mu \nn\\
&+ \hn \cdot \hat j + (B_\mu + W_\mu) j_\mu + \X_\mu \cdot \vec j_\mu. 
\eea
Notice that here we have inserted the delta-function 
gauge condition $\delta (W_\mu-W^{(0)}_\mu)$
to fix the the gauge degrees of freedom created by
the decomposition $A_\mu= B_\mu+W_\mu$.
In principle here $W^{(0)}_\mu$ can be any constant field, but we 
might choose $W^{(0)}_\mu=0$ for simplicity. With this understanding
the above generating functional reduces to
\bea
&W\{\hat j, j_\mu, \vec j_\mu\} = \dfrac{}{}\int {\cal D} \hn {\cal D}
B_\mu {\cal D} \X_\mu {\cal D} \vec{\eta} ~{\cal D}\vec{\eta}^{~*}
{\cal D} \vec{c} ~{\cal D}\vec{c}^{~*} \exp[\dfrac{}{}
i\int{\cal L}_{eff} d^4x], \nn\\
&{\cal L}_{eff} =-\dfrac{1}{4}
{\check F}_{\mu\nu}^2 
-\dfrac{1}{2\xi_1}(\partial_\mu \check A_\mu)^2
+ {\vec c}^{~*}\partial_\mu \check D_\mu \vec c + \lambda (\hn^2-1) \nn\\ 
&-\dfrac{1}{4} ( \check D_\mu \X_\nu -
\check D_\nu \X_\mu)^2  
-\dfrac{g}{2} {\check F}_{\mu\nu}
\cdot (\X_\mu \times \X_\nu) 
- \dfrac{g^2}{4} (\X_\mu \times \X_\nu)^2 \nn\\
&- \dfrac{1}{2 {\xi}_2} (\check D_\mu {\X}_\mu)^2
+ {\vec \eta}^{~*}\check D_\mu D_\mu \vec \eta 
+ \lambda_\mu \hat n \cdot \X_\mu \nn\\
&+ \hn \cdot \hat j + B_\mu j_\mu + \X_\mu \cdot \vec j_\mu.
\eea
The difference in the two schemes is that in the second scheme 
the ghost fields $\vec \eta$ and ${\vec \eta}^{~*}$ become isotriplets,
and the ghost interaction of the passive type becomes simpler.
This is because here $\check D_\mu \W_\mu$ need no longer 
be orthogonal to $\hn$.

Observe that, with the identification of $B_\mu$ as $A_\mu$,
the effective Lagrangian (36) is what we would have
obtained with the previous quantization, had we identified the
passive gauge transformation by 
\bea
\delta \hat A_\mu =0, ~~~~~~~\delta \X_\mu 
= \dfrac {1}{g} \hat D_\mu \alpha,
\eea
but not by (23). Strictly speaking, this identification is inconsistent with
the definition $A_\mu = \hn \cdot \vec A_\mu$ and (22). This was why we did not 
adopt this gauge transformation in our 
previous quantization. But the lesson that
we learn from the above analysis is that the above passive gauge 
transformation (37) could actually be acceptable and justifiable after all.

To be complete we will now discuss an alternative (a third) quantization.
Remember that the new gauge degrees of freedom originates from
the arbitrariness of the decomposition for a fixed $\vec A_\mu$.
Since the theory must be independent of the decomposition, 
we could quantize the theory just like in the perturbative QCD 
with the gauge condition 
\bea
\partial_\mu \vec A_\mu = \partial_\mu \hat A_\mu 
+ \partial_\mu \X_\mu =0, 
\eea
and take into account the arbitrariness of the decomposition with the 
following mathematical identity,
\bea
\int {\cal D} \hn ~{\rm det}  {\Big (}\dfrac {\delta 
(\hat D_\mu \vec X_\mu)}{\delta \hn} {\Big )}
~\delta (\hat D_\mu \vec X_\mu) = 1.
\eea
For a fixed $\vec A_\mu$ we can calculate the determinant with the observation 
\bea
J_{ab} = \dfrac {\delta (\hat D_\mu \vec X_\mu)_a}{\delta n^b}
= - \dfrac {(D_\mu \delta \hat A_\mu)_a}{\delta n^b}.
\eea
But an important point here is that we must take into account
the following constraints
\bea
\delta (\hn \cdot \hat D_\mu \vec X_\mu) 
= \hn \cdot \delta (\hat D_\mu \vec X_\mu) = 0,
~~~~~~~\hn \cdot \delta \hn = 0,
\eea
in the evaluation of the determinant, which make 
\bea
n_a J_{ab} = J_{ab} n_b = 0.
\eea
So, here again the ghost fields corresponding to the determinant
should be orthogonal to $\hn$ and have only two degrees.
Now, the Lorentz gauge (38) is a perfect gauge for us, but
here we will in stead choose a slightly different gauge
condition
\bea
\hat D_\mu \vec A_\mu = \partial_\mu \hat A_\mu 
+ \hat D_\mu \X_\mu =0,
\eea
for the purpose of comparison with the previous quantization 
schemes. In this case the corresponding determinant is given by
\bea
M_{ab} = \dfrac {\delta (\hat D_\mu \vec A_\mu)_a}{\delta \alpha^b}
= (\hat D_\mu D_\mu)_{ab}.
\eea
With this we obtain the following generating functional
\bea
&W\{\hat j, j_\mu, \vec j_\mu\} = \dfrac{}{}\int {\cal D} \hn {\cal D}
A_\mu {\cal D} \X_\mu {\cal D} \vec{\eta_\perp} ~{\cal D}\vec{\eta_\perp}^{*}
{\cal D} \vec{c} ~{\cal D}\vec{c}^{~*} \exp[\dfrac{}{}
i\int{\cal L}_{eff} d^4x], \nn\\
&{\cal L}_{eff} =-\dfrac{1}{4}
{\hat F}_{\mu\nu}^2 
-\dfrac{1}{4} ( \D_\mu \X_\nu -
\D_\nu \X_\mu)^2 
-\dfrac{g}{2} {\hat F}_{\mu\nu}
\cdot (\X_\mu \times \X_\nu) 
- \dfrac{g^2}{4} (\X_\mu \times \X_\nu)^2 \nn\\ 
&-\dfrac{1}{2\xi_1}(\partial_\mu \hat A_\mu)^2 
+ {\vec c}^{~*}\hat D_\mu D_\mu \vec c 
- \dfrac{1}{2 {\xi}_2} (\hD_\mu {\X}_\mu)^2 
+ {\vec \eta_\perp}^{~*} \cdot (\partial_\mu \hn + g\X_\mu \times \hn)
(\hat A_\mu + \X_\mu) \cdot \vec \eta_\perp \nn\\
&+ {\vec \eta_\perp}^{~*} \cdot A_\mu (\partial_\mu + \hat D_\mu) 
\vec \eta_\perp -(\X_\mu \cdot \partial_\mu \hn) 
({\vec \eta_\perp}^{~*} \cdot \vec \eta_\perp) 
+ \dfrac {1}{g} {\vec \eta_\perp}^{~*} \cdot 
(\partial^2 \hn \times \vec \eta_\perp
- \hn \times \partial^2 \vec \eta_\perp) \nn\\
&+ \lambda (\hn^2-1) + \lambda_\mu \hat n \cdot \X_\mu 
+ \hn \cdot \hat j + A_\mu j_\mu + \X_\mu \cdot \vec j_\mu.
\eea
Again notice that, although the ghost interaction of $\vec \eta_\perp$ 
and ${\vec \eta_\perp}^{~*}$ is 
introduced to remove the extra gauge 
degrees of freedom caused by different choices of $\hn$, it does not
remove the topological field $\hn$ from the theory.
Obviously the three quantization schemes are based on different 
choices of decomposition. But from the physical point of view 
they should be equivalent to each other, and thus describe 
the same non-Abelian dynamics. It would be very nice to have
a rigorous proof of the equivalence of the above three quantization schemes. 

The modified theory obtained with the decomposition (1)
has been called the restricted QCD (without $\X_\mu$), 
or the extended QCD (with $\X_\mu$) \cite{cho1,cho2}.  We 
emphasize that, even without $\X_\mu$, the restricted QCD
described by
\bea
{\cal L}_{eff} =-\dfrac{1}{4}
{\hat F}_{\mu\nu}^2 
-\dfrac{1}{2\xi}(\partial_\mu \hat A_\mu)^2
+ {\vec c}^{~*}\partial_\mu \hat D_\mu \vec c + \lambda (\hn^2-1),
\eea
makes a non-trivial self-consistent theory. It has a full $SU(2)$
gauge degrees of freedom with the non-Abelian monopole
as an essential ingredient, and describes a very interesting
dual dynamics of its own. 

The serious question now is whether this modification 
induced by the decomposition (1) describes the same
physics or not. We believe that this is so. {\it In fact, we believe that
the conventional QCD makes sense only when one expands 
it perturbatively around the trivial vacuum, but 
becomes incomplete otherwise in the sense that it does not
properly take into account its topological structures}. For
instance, it can not define the topological charge of the Wu-Yang
monopole, in spite of the fact that the theory obviously has the
topological monopole. Moreover it has never been able to define the color
direction in a gauge independent way. In other words, it has 
never been able to define the conserved color charge which is 
gauge invariant which is supposed to be confined. In comparison in our
extended QCD the topological field $\hat n$ allows us to define
not only the monopole charge with the mapping $\pi_2 (S^2)$, but
also the gauge independent color direction and gauge invariant
color charge uniquely. In fact we can easily obtain the gauge invariant
conserved color charge from our equation of motion (12).
For these reasons
we believe that only our extended QCD is able to describe the
dynamics of the non-Abelian gauge theory unambiguously.

Before we leave this section it is worth to remark that
our decomposition (1) can actually
``Abelianize'' (or more precisely ``dualize'') the non-Abelian
dynamics. To see this let
\begin{eqnarray}
&\vec{X}_\mu =X^1_\mu ~\hat{n}_1 + X^2_\mu ~\hat{n}_2\mbox{,} \nn\\
&(X^1_\mu = \hat {n}_1 \cdot \vec X_\mu,~~~X^2_\mu =
\hat {n}_2 \cdot \vec X_\mu)            \nonumber
\end{eqnarray}
and find
\begin{eqnarray}\label{b2}
\hat{D}_\mu \vec{X}_\nu &=&[\partial_\mu X^1_\nu-g
(A_\mu+ \tilde C_\mu)X^2_\nu]~\hat n_1
+ [\partial_\mu X^2_\nu+ g (A_\mu+ \tilde C_\mu)X^1_\nu]~\hat{n}_2.
\end{eqnarray}
So with
\bea
{\cal A}_\mu = A_\mu + \tC_\mu,
~~~~~X_\mu = \dfrac{1}{\sqrt{2}} ( X^1_\mu + i X^2_\mu ),
\eea
one could express the Lagrangian explicitly in terms of the dual
potential $B_\mu$ and the complex vector field $X_\mu$,
\begin{eqnarray}
&{\cal L}=-\frac{1}{4}(F_{\mu\nu}+ H_{\mu\nu})^2
-\frac{1}{2}|\hat{D}_\mu{X}_\nu-\hat{D}_\nu{X}_\mu|^2 \nn\\
&+ ig (F_{\mu\nu} + H_{\mu\nu}) X_\mu^* X_\nu
-\dfrac{1}{2} g^2 [(X_\mu^*X_\mu)^2-(X_\mu^*)^2 (X_\nu)^2],
\end{eqnarray}
where now
\bea
\hat{D}_\mu{X}_\nu = (\partial_\mu + ig {\cal A}_\mu) X_\nu.  \nonumber
\eea
Clearly this describes an Abelian gauge theory coupled to
the charged vector field $X_\mu$.
But the important point here is that the Abelian potential
${\cal A}_\mu$ is given by the sum of the electric 
and magnetic potentials $A_\mu+
\tilde C_\mu$.
In this Abelian form the equation of motion (12) is re-expressed as
\begin{eqnarray}
&\partial_\mu(F_{\mu\nu}+ H_{\mu\nu}+X_{\mu\nu}) = i g [X^*_\mu
(\hat D_\mu X_\nu - \hat D_\nu X_\mu) - X_\mu (\hat D_\mu X_\nu - \hat D_\nu
X_\mu )^*], \nonumber
\\
&\hat{D}_\mu(\hat{D}_\mu X_\nu- \hat{D}_\nu X_\mu) = ig X_\mu
(F_{\mu\nu}+H_{\mu\nu} +X_{\mu\nu}),
\end{eqnarray}
where now
\bea
X_{\mu\nu} = - i g ( X_\mu^* X_\nu - X_\nu^* X_\mu ).  \nonumber
\eea
This shows that one can indeed Abelianize the non-Abelian theory
with our decomposition. But notice that here we have never fixed
the gauge to obtain this Abelian formalism, and one might
ask how the non-Abelian gauge symmetry is realized in this ``Abelian''
theory. To discuss this let
\bea
&\vec \alpha = \alpha_1~\hn_1 + \alpha_2~\hn_2 + \theta~\hn, 
~~~~~~~\vec C_\mu = - \dfrac {1}{g} \hn \times \partial_\mu \hn
= - C^1_\mu \hn_1 - C^2_\mu \hn_2, \nn\\
&\alpha = \dfrac{1}{\sqrt 2} (\alpha_1 + i ~\alpha_2),
~~~~~~~C_\mu = \dfrac{1}{\sqrt 2} (C^1_\mu + i ~ C^2_\mu).
\eea
Then the Lagrangian (49) is invariant not only under 
the active gauge transformation (3) described by 
\bea
&\delta A_\mu = \dfrac{1}{g} \partial_\mu \theta - 
i (C_\mu^* \alpha - C_\mu \alpha^*), 
~~~~~~~&\delta \tilde C_\mu = - \delta A_\mu, \nn\\
&\delta X_\mu = 0,
\eea
but also under the passive gauge transformation (23) described by
\bea
&\delta A_\mu = \dfrac{1}{g} \partial_\mu \theta - 
i (X_\mu^* \alpha - X_\mu \alpha^*), ~~~~~~~&\delta \tilde C_\mu = 0, \nn\\
&\delta X_\mu = \oneg \hD_\mu \alpha - i \theta X_\mu.
\eea
This tells that the ``Abelian'' theory not only retains 
the original gauge symmetry, but actually has an enlarged (both the active
and passive) gauge symmetries. So the only change in this ``Abelian''
formulation is that here the topological field $\hn$ is
replaced by the magnetic potential $\tilde C_\mu$.
But we emphasize that this is not the ``naive'' Abelianization
of the $SU(2)$ gauge theory.
The difference is that here the Abelian gauge
group is actually made of $U(1)_e \otimes U(1)_m$, so that
the theory becomes a dual gauge theory where the magnetic
potential plays the crucial role \cite{cho1,cho2}.
This must be obvious from (52) and (53).

\section{Comparison between QCD and Skyrme-Faddeev Theory}

Now, we review the physical content of the Skyrme-Faddeev theory
and discuss the similarities between the Skyrme-Faddeev theory and
QCD. To do this we start from the Skyrme-Faddeev Lagrangian 
\bea
{\cal L}_{SF} = - \dfrac{\mu^2}{2} (\partial_\mu \hat n)^2 -
\dfrac{g}{4} (\partial_\mu \hat n \times \partial_\nu \hat n)^2,
\eea 
which gives the following equation of motion 
\bea \hn \times
\partial^2 \hn - \dfrac{g^2}{\mu^2} ( \partial_\mu H_{\mu\nu} )
\partial_\nu \hn = 0. 
\eea 
This allows the Faddeev-Niemi knot
solutions \cite{faddeev1,faddeev2}. But obviously the knots are
the solitons, and probably can not describe the elementary object
of the theory. To find the elementary ingredient of the theory,
notice that the equation of motion (55) has another solution which
is much simpler. Let the polar coodinates of $R^3$ be
$(r,~\theta,~\phi)$ and let
\bea 
\hat n = \hat r = (\sin{\theta} \cos{\phi}, ~\sin{\theta} \sin{\phi},
~\cos{\theta}). 
\eea 
Clearly we have
\bea
\partial^2 \hat r = - \dfrac {2}{r^2} \hat r, ~~~\partial_\mu H_{\mu\nu} =0,
\eea
so that (56) becomes a solution of (55), except at the origin.
Furthermore, the magnetic field it creates is identical to that of the
Wu-Yang monopole in the $SU(2)$ gauge theory sitting at
the origin. Of course this monopole is also topological, whose
topological quantum number is given exactly by the same mapping
$\pi_2 (S^2)$ defined by $\hat n$ as the Wu-Yang monopole. 
{\it This means that the
non-linear sigma field $\hat n$ in the Skyrme-Faddeev theory
describes the non-Abelian monopole, and the Faddeev-Niemi knots
are the magnetic flux tubes made of the monopole-anti-monopole
pair. The interesting point here is that, unlike in the Abelian
theory, these flux tubes are able to form the topologically stable
knots due to the non-linear self-interaction}. 
This tells that both the non-Abelian gauge theory and the
Skyrme-Faddeev theory describe a non-trivial monopole dynamics.
Notice that (56) forms a solution of (54) even without the
non-linear interaction. {\it This means that the non-linear sigma model
in general should really be viewed as a theory of monopole, even
without the non-linear self interaction}.

With this we are ready to prove that the Faddeev-Niemi knots can
actually describe the multiple vacua of the $SU(2)$ QCD. 
To show that the knot solutions of (55) can indeed be used to
describe the vacuum solutions of our equation of motion (12),
notice first that any $\hat n$
which describes the Faddeev-Niemi knots is smooth everywhere in
$R^3$, and thus defines the mapping $\pi_3 (S^2)$.
Secondly, for any knot described by (55) we can always
introduce the magnetic potential $\tilde C_\mu$ 
which is smooth everywhere in $R^3$ through (5). Let
$\tilde C_\mu$ be the magnetic potential of the 
knot described by $\hat n$, 
\bea 
H_{\mu\nu} = \partial_\mu \tilde C_\nu - \partial_\nu \tilde C_\mu 
= -\dfrac{1}{g} \hat{n}\cdot(\partial_\mu\hat{n}\times\partial_\nu\hat{n}).
\nonumber
\eea
Then the knot quantum number $k$ of $\pi_3 (S^2)$ is given 
by \cite{faddeev1,white} 
\bea
k = \dfrac{g^2}{32\pi^2} \int \epsilon_{ijk} \tilde C_i H_{jk} d^3 x.
~~~~~~(i,j,k = 1,2,3) 
\eea
Now, consider the following static potential in $SU(2)$ QCD which is
smooth everywhere in $R^3$,
\bea 
\hat A_\mu
= - \tC_\mu \hn - \oneg \hn \times \partial_\mu \hn, ~~~ \vX_\mu = 0, 
\eea 
where $\tC_\mu$ and $\hn$ are given by the knot solution.
It must be clear from (5) that this potential 
produces a vanishing field strength, 
and thus forms a vacuum solution
of our equation of motion (12). Furthermore, in this case 
the vacuum quantum number m
of the potential (59) which describes $\pi_3 (S^3)$ 
is defined by \cite{bpst,thooft}
\bea
m = \dfrac{g^2}{96\pi^2} \int \epsilon_{ijk} 
\vA_i \cdot (\vA_j \times \vA_k) d^3 x.
\eea
But this is exactly the same mapping which defines the knot
quantum number (58) with the Hopf fibering \cite{cho4}. 
In fact, with (59) one can easily show that (60) reduces to
(58). {\it This proves that indeed the
Faddeev-Niemi knots can be viewed to describe the multiple vacua of $SU(2)$ QCD,
and the topological quantum number of the knots becomes nothing but
the vacuum quantum number}. Actually we can simply claim that the
knots describe the multiple vacua of the restricted QCD, because
(59) obviously forms the vacuum solutions
of the restricted QCD. We emphasize that the vacuum solutions (59)
become nothing but the multiple vacuum solutions of a spontaneously 
broken $SU(2)$ gauge theory that we proposed long time ago, 
if we identify $\hn$ as the normalized isotriplet
(Higgs) scalar field $\hat \phi$ in the spontaneously broken
theory \cite{cho4}. In this sense one can also
claim that the Faddeev-Niemi knots also describe the multiple vacua of 
the spontaneously broken $SU(2)$ gauge theory.

We have shown that the non-linear sigma model in general is a theory of 
monopoles. An important question then is what are the physical states 
of the Skyrme-Faddeev theory. Most probably the monopole is not 
likely to be a physical state, because it has an infinite energy
(Remember here that the monopole solution (56) describes a classical state, not
a quantum state). {\it This tells that the Skyrme-Faddeev theory, just like 
QCD, is probably a theory of confinement, in which the non-linear
self interaction of the monopoles confines the long range magnetic flux
of the monopoles}. In this view the Faddeev-Niemi knots can
be viewed as the confined ``glueball states'' of the non-linear
sigma model which are 
made of the monopole-anti-monopole pairs. 
The new feature here is that these ``glueball states'' (unlike those in
QCD) have the topological quantum number which makes them 
topologically stable.

\section{Discussion}

The existence of the gauge independent reparametrization (1)
of the non-Abelian potential in terms of the restricted  
potential $\hat {A}_{\mu}$ and the
valence potential $\vec {X}_{\mu}$
has been known for more than twenty years \cite {cho1,cho2},
but its physical significance
have become appreciated only recently \cite{faddeev2,lang,gies}.
In this paper we have discussed the physical impacts of
our reparametrization (1)  
on QCD. We have discussed how the reparametrization
modifies the non-Abelian dynamics, and presented three quantization schemes of
the extended QCD. In particular, we have shown that the reparametrization 
enlarges the dynamical degrees of QCD by making 
the topological field $\hn$ dynamical upon quantization. Furthermore,
with the reparametrization, we have demonstrated 
that the Skyrme-Faddeev theory of
non-linear sigma model and $SU(2)$ QCD have almost identical
topological structures. In particular we have shown that both 
are not only theories of monopoles but also theories of 
confinement, where the 
monopole plays a crucial role in their dynamics.  
Together with the Faddeev-Niemi
conjecture the parallel between the two theories is indeed
remarkable.

A straightforward consequence of our decomposition is the
existence of the restricted QCD described by (49). What is remarkable 
with the restricted theory is that we can construct a non-Abelian gauge theory
with much simpler gauge potential (i.e., with the
restricted gauge potential), which nevertheless has the full topological 
characters of the non-Abelian gauge symmetry. The theory
contains the non-Abelian monopole as an essential ingredient,
and describes a very interesting 
dual dynamics of its own. So it must be obvious that the
restricted theory should play a fundamental role in QCD.
Furthermore we can construct a whole family of interesting new non-Abelian
gauge theories by adding any gauge-covariant
colored source to the restricted gauge theory.
  
We conclude with the following remarks. \\
1)  Our analysis clarifies the physical
content of the Skyrme-Faddeev theory. Indeed
the physical content of the Skyrme-Faddeev theory of non-linear
sigma model has never been clear. Faddeev and Niemi have shown
that the theory is probably a theory of knots. But our analysis
tells that it is a self-interacting theory of monopole. 
The interesting point here is that the non-linear
interaction can actually screen (i.e., confine) the magnetic
flux of the monopole, and force it to form a flux tube.
Moreover, the topology of the theory can make the flux to
form the stable knots. But what is really surprising
is that, in spite of the fact that the Skyrme-Faddeev theory
contains a non-trivial mass parameter, it admits a long range
monopole solution. The elementary particle of the theory
is the massless monopole, and the mass scale of the theory
describes (not the mass of the monopole but) the penetration scale
(i.e., the confinement scale) 
of the magnetic flux made of the monopole-anti-monopole pair.
This is really remarkable. \\
2) Although the similarities between the Skyrme-Faddeev theory
and QCD is striking, it should be emphasized that they are really
different from the physical point of view. 
For example, the Faddeev-Niemi conjecture suggests that the knots
might be interpreted to describe the glueball states
in QCD \cite{faddeev2,shab}.
This is an overstatement. Although the two theories are the theories 
of confinement where the monopole dynamics plays the important
role, their confinement mechanism is really different.
In QCD the monopole condensation generates the confinement, 
but in Skyrme-Faddeev theory it is the self-interaction of the monopoles
which creates the confinement. In this respect we emphasize that 
the Skyrme-Faddeev theory (unlike QCD) already has the confinement
scale, and does not need any dynamical symmetry breaking to generate
the confinement. Furthermore, the Faddeev-Niemi knots
are the magnetic knots. But in QCD the gluons could form only
the electric flux tubes after the monopole condensation. 
This means that, at the best,
the Faddeev-Niemi knots can provide a dual description of the
possible electric knots. If this is so, the really interesting
question is whether QCD can actually allow 
any electric knot which is stable.
Although this is an interesting issue worth further investigation,
it would be difficult to establish the topological stability for
the glueballs in QCD \cite{fadd3}. \\
3) We hope that our analysis completely settles the controversy around
the question whether one need any extra constraint 
to kill the topological degrees of $\hn$ with our
decomposition (1) \cite{shab,gies}. The condition (15) is 
nothing but a gauge condition
which one need to remove the unphysical degrees 
of $\vX_\mu$ to keep it massless,
which has nothing to do with the topological degrees of $\hn$. 
Furthermore, even with (15), the topological field $\hn$ becomes
dynamical after the gauge fixing. As we have emphasized, any constraint
for $\vec A_\mu$ (and $\vX_\mu$) should come from a gauge fixing,
and we have shown how the condition (15) 
really becomes a gauge condition, especially in the quantization
by the background field method.
In fact with this gauge condition one could successfully calculate
the effective action of QCD in the background field method. 
Remarkably the resulting effective action allows us to establish
the monopole condensation at one loop level, and to
demonstrate that the monopole
condensation is the true vacuum of QCD \cite{cho3,cho7}.
Most importantly, the topological field $\hn$ already plays 
a fundamental role in the restricted QCD even without $\vX_\mu$. 
This tells that any attempt to remove the topological field imposing
a constraint on $\vX_\mu$ is futile.

Note Added: The fact that the Faddeev-Niemi knots could
be interpreted to describe the multiple vacua in $SU(2)$
QCD has also been re-discovered recently by van Baal and 
Wipf, who obtained an identical result in a different context.
See P. van Baal and A. Wipf, hep-th/0105141, Phys. Lett. {\bf B},
in press. 

\vspace{0.5cm}\hspace{-0.5cm}
{\bf Acknowledgements}\par
\vspace{0.5cm}
One of the authors (YMC) thanks L. Faddeev, and A. Niemi
for the fruitful discussions, and Professor C. N. Yang for
the continuous encouragements.
The work is supported in part by Korea Research Foundation (Grant KRF-2000
-015-BP0072) and by the BK21 project of Ministry of Education.


\end{document}